\documentstyle[11pt]{article}

\begin{document}

\title{\bf Spin Wave Specific Heat in Quasiperiodic Fibonacci Structures}

\author{C.G. Bezerra$^{\rm a}$, E.L. Albuquerque$^{\rm b}$ \thanks
{Corresponding author, e-mail: ela@dfte.ufrn.br}, A.M. Mariz$^{\rm
b}$,\\ L.R. da Silva$^{\rm b,c}$, and C. Tsallis$^{\rm d}$}
\maketitle

\noindent (a) Department of Physics and Astronomy, University of
Western Ontario, London-Ontario N6A3K7, Canada
 \vskip 0.1 cm \noindent
(b) Departamento de F\'\i sica, Universidade Federal do Rio Grande
do Norte, Natal-RN 59072-970, Brazil
 \vskip 0.1 cm \noindent
(c) Department of Physics, Boston University, Boston-MA 02215, USA
\vskip 0.1 cm \noindent
(d) Centro Brasileiro de Pesquisas
F\'{\i}sicas, Rua Xavier Sigaud 150, 22290-180 Rio de Janeiro-RJ,
Brazil

\begin{abstract}
The energy spectra of a variety of collective modes on
quasiperiodic structures exhibit a {\it complex fractal profile}.
Among the modes that have attracted particular attention in this
context, are the spin wave spectra of quasiperiodic magnetic
multilayers that obey a substitutional sequence of the Fibonacci
type. They are described within the framework of the Heisenberg
theory. In order to have a deep insight on the relevant
thermodynamical implications of the above mentioned energy
spectra's fractal profile, we have performed analytical and
numerical calculations of the spin wave specific heat associated
with successive hierarchical sequences of the Fibonacci
quasiperiodic structures. The spectra show interesting oscillatory
behavior in the low-temperature region, which can be traced back
to the spin wave's self-similar energy spectrum.
\end{abstract}

\vskip 0.2 cm \noindent {\it PACS:} 05.20-y; 61.43.Hv; 61.44.Br;
75.30.Ds

\noindent {\it Keywords:} Quasi-Crystals; Spin waves; Fractal
behavior; Thermodynamical properties

\newpage

\section{Introduction}
Since the discovery of the icosahedral phase in Al-Mn alloys by
means of X-ray spectroscopy \cite{1}, the quasicrystaline systems
have been extensively studied (for a revision see \cite{2}). In
particular, the physical properties of the so-called quasiperiodic
structures have attracted a lot of attention from both theoretical
and experimental point of view. Although the term {\it
quasicrystal} is more appropriate for natural compounds or
artificial alloys,  in one dimension there is no difference
between this case and the quasiperiodic structure formed by the
incommensurate arrangement of periodic unit cells. Due to this
motivation,  Merlin and collaborators,   using the molecular beam
epitaxy (MBE) technique, grew in 1985 the first quasiperiodic
superlattice following the Fibonacci sequence \cite{3}. After that
pioneer work, other quasiperiodic structures were experimentally
realized \cite{4}.

>From a theoretical point of view, a number of physical properties
has been studied in quasiperiodic structures. Among them we can
cite the energy spectra of polaritons \cite{5}, phonons \cite{6},
electrons \cite{7} and spin waves \cite{8,9}, as well as the
magnetoresistance and magnetization curves of quasiperiodic thin
films \cite{10}. A quite interesting feature, common to all of
these systems, is a self-similar pattern of their spectra. In
fact, the energy spectra of the above referred particles and
systems are highly fragmented and tend to Cantor sets in the
thermodynamic limit. The origin of this fractality can be
attributed to the long range order induced by the non-usual
hierarchical structure of the quasiperiodic sequences used in the
construction of the system.

A very interesting question aroused in the last few years: {\it
what are the consequences of a fractal energy spectra on the
behavior of the thermodynamic properties of quasiperiodic
systems?} In a recent work, Tsallis and collaborators \cite{11}
studied the specific heat properties of a fractal energy spectra
generated by the {\it geometrical triadic Cantor set}. They have
found, as the main result of the paper, that the specific heat
presents oscillations around the fractal dimension of the spectra.
In addition, a non-uniform convergence between the so-called
banded and discrete models was observed. Their results were
extended by Vallejos {\it et al} \cite{12,13} for the {\it
two-scale} Cantor set case. For this more general situation, the
specific heat also exhibits log-periodic oscillations around the
fractal dimension. Later Carpena {\it et al} \cite{14}, using the
properties of a {\it multifractal} spectra showed under what
conditions the oscillatory regime disappears. Finally, Curado and
Rego-Monteiro \cite{15} examined the thermodynamic properties of a
solid exhibiting the energy spectrum given by a logistic map.

The aim of this work is to push a little bit more the
understanding of the thermodynamic properties of quasicrystals,
analyzing  the specific heat of a {\it real} fractal spectra of
spin waves in quasiperiodic Fibonacci magnetic superlattices. We
consider that the building blocks used here to set up the
quasiperiodic structures are ferromagnetic materials, whose
dynamics are described by the Heisenberg Hamiltonian. Our main
concern is to emphasize the differences between the specific heat
profile obtained by using a {\it real} spin wave's {\it
multifractal} energy spectra for a given in-plane wavevector,
(with, by instance, their proper scaling laws), as discussed in
Ref. 8, and an {\it idealized} Cantor set, considered in previous
works. Throughout the paper we use the classical Maxwell-Boltzmann
statistics.

This paper is organized as follows: In section II we discuss the
theoretical method used to obtain the spin waves dispersion
relation (their energy spectra), and the rules used to build up
the Fibonacci superlattice. Section III is devoted to the specific
heat theory of a general scaled energy band spectra. The
application of this theory to the Fibonacci quasiperiodic
structure is then made, and the main features of the spectra
obtained are discussed. Finally, section IV presents the
conclusions of this work.

\section{Physical model}
In this section we briefly describe the physical model for the
quasiperiodic magnetic superlattices. A detailed description can
be found in Ref. \cite{8}. We consider superlattices in which
$n_{A}$ layers of material $A$ (building block $A$) alternate with
$n_{B}$ layers of material $B$ (building block $B$). Both
materials are taken to be simple cubic spin-$S$ Heisenberg
ferromagnets having exchange constants $J_{A}$ and $J_{B}$,
respectively, and lattice constant $a$. At the interface $A/B$,
the exchange constant is suppose to be equal to $I$. The
Heisenberg Hamiltonian for each component is:

\begin{equation}
{\cal H}=(-1/2){\sum\limits_{i,j}{J_{ij} \stackrel{\rightarrow
}{S}_i\cdot \stackrel{\rightarrow }{S}_j}}-g\mu
_BH_0{\sum\limits_i{S_i^z}}.
\end{equation}

Here the sum in the first term is over sites $i$ and nearest
neighbors ($n.n.$) $j$, and $H_{0}$ is a static external magnetic
field pointing in the $z$-direction .

The spin wave dispersion relation in a superlattice can be found
by solving the equations of motion for the operator
$S_i^{+}=S_i^x+iS_i^y$, i.e:

\begin{equation}
\hbar \frac \partial {\partial t}S_i^{+}=g\mu
_BH_0S_i^{+}+<S^z>\sum\limits_{n.n.}J_A(S_i^{+}-S_j^{+}),
\end{equation}

\noindent where $<S^z>$ is the random phase approximation for the
$z$- component of the spin operator. The solution of (2), for
material $A$, is:

\begin{equation}
S_i^{+}=[ A_l\exp (i\stackrel{\rightarrow }{k}_A\cdot \stackrel{
\rightarrow }{r}_A) + B_l\exp (-i\stackrel{ \rightarrow
}{k}_A\cdot \stackrel{\rightarrow }{r}_A)] (\exp -i\omega t)
\end{equation}

\noindent with similar expression for  material $B$. These
solutions are linked together using the equation of motion (2) at
the boundaries of the $n$-th unit cell, which can be written in a
matricial form. Using the translational invariance property of the
excitations, through Bloch's theorem, the spin wave dispersion
relation follows as:
\begin{equation}
\cos (QD)=(1/2)Tr\left[ {\mathbf T}\right].
\end{equation}
Here $Q$ is the Bloch'wavevector of the collective mode, $D$ is
the size of the superlattice's unit cell, and ${\mathbf T}$ is a
transfer matrix which relates the coefficients of the $(l+1)th$
cell to the coefficients of the preceding one.

A  Fibonacci superlattice can be grown experimentally by
juxtaposing the two building blocks $A$ and $B$ in such way that
the $nth$-generation of the superlattice $S_{n}$ is given
iteratively by the rule $S_{n}=S_{n-1}S_{n-2}$, for $n\geq2$, with
$S_{0}=B$ and $S_{1}=A$. It is also invariant under the
transformations $A\rightarrow AB$ and $B\rightarrow A$. The
Fibonacci generations are,
\begin{equation}
S_{0}=[B], \hspace{0.1cm} S_{1}=[A], \hspace{0.1cm} S_{2}=[AB],
\hspace{0.1cm} S_{3}=[ABA], \hspace{0.1cm} etc.
\end{equation}
The number of building blocks increases according to the Fibonacci
number, $F_{n}=F_{n-1}+F_{n-2}$ (with $F_{0}=F_{1}=1$), and the
ratio between the number of the building blocks $A$ and the number
of building blocks $B$, when $n>>1$, in the sequence tends to
$\tau = (1/2) (1+\sqrt{5})$, an irrational number known as the
golden mean.

It can be shown that the transfer matrix for the $nth$ generation
of a Fibonacci superlattice can be obtained by a simple recurrence
relation given by \cite{8},
\begin{equation}
{T_{s_{n}}}= {T_{s_{n-2}}} \cdot {T_{s_{n-1}}}, \quad n\geq2.
\end{equation}
Therefore, from the knowledge of the transfer matrices $T_{S_{0}}$
and $T_{S_{1}}$, we can determine the transfer matrix for any
generation and consequently the dispersion relation. Note that the
matrix $T_{S_{2}}$ recovers the periodic case.

\section{Specific heat Spectra}
The spin wave fractal spectra for the  Fibonacci superlattices is
depicted in Fig.\ 1, for a fixed value of the in-plane
dimensionless wavevector $k_{x}a$. From there we can see the
forbidden and allowed energies  of the spin wave spectra against
the  Fibonacci's generation number $n$, up to their $8th$
generation, which corresponds an unit cell with $21$ $A$'s and
$13$ $B$'s building blocks. The number of allowed bands is equal
to three times the Fibonacci number $F_{n}$, of the correspondent
generation. Notice that, as expected, for large $n$ the allowed
band regions get narrower and narrower and they have a typical
Cantor set structure.

We address now to the specific heat of the spectra indicated in
Fig.\ 1. The description below is general and can be applied to
any banded spectrum. In Fig.\ 1 each spectrum, for a fixed
generation number $n$, has $m$ allowed continuous bands. We
consider, without loss of generality, the level density within
each band constant. The partition function for the $nth$
generation is then given by:

\begin{equation}
Z_{n}=\int_{0}^{\infty}\rho (\epsilon)e^{-\beta
\epsilon}d{\epsilon},
\end{equation}
Here $\beta=1/T$ (by choosing the Boltzmann's constant $k_{B}=1$),
and we take the density of states $\rho (\epsilon)=1$. After a
straightforward calculation we can write $Z_{n}$ as,

\begin{equation}
Z_{n}={\frac {1}{\beta}}\sum_{i=1,3,\ldots}^{2m-1}
e^{-\beta\epsilon_{i}}[1-e^{-\beta\Delta_{i}}].
\end{equation}
Here the subscript $n$ is the generation number, $m$ is the number
of allowed bands and $\Delta_{i}=\epsilon_{i+1}-\epsilon_{i}$ is
the difference between the top and bottom energy levels of each
band.

The specific heat is then given by,
\begin{equation}
C_{n}(T)={\frac {\partial}{\partial T}}[T^2{\frac {\partial \ln
Z_{n}}{\partial T}}],
\end{equation}
which can be written as,
\begin{equation}
C_{n}(T)=1+{\frac {\beta f_{n}}{Z_{n}}}-{\frac
{g^2_{n}}{Z^2_{n}}},
\end{equation}
with,
\begin{equation}
f_{n}=\sum_{i=1,3,\ldots}^{2m-1}[ \epsilon ^2_{i}
e^{-\beta\epsilon_{i}}-\epsilon ^2_{i+1}e^{-\beta\epsilon_{i+1}}].
\end{equation}
and,
\begin{equation}
g_{n}=\sum_{i=1,3,\ldots}^{2m-1}[ \epsilon_{i}
e^{-\beta\epsilon_{i}}-\epsilon_{i+1}e^{-\beta\epsilon_{i+1}}].
\end{equation}

Therefore, once we know the energy spectra of the spin wave which
propagates in a given sequence's generation of a quasiperiodic
structure, we can determine the associated specific heat's spectra
by using (10).

Fig.\ 2  shows the spin wave specific heat spectra of the
Fibonacci superlattices, for the in-plane wavevector $k_{x}a=2.0$,
as a function  of the temperature. For the high temperature limit
($T\rightarrow\infty$), the specific heat  for all generation
numbers converges and decays as $T^{-2}$, for arbitrary $n$, in
agreement with the triadic case. This is a consequence of the
existence of a maximum energy value in the spectrum (once the
spectrum is bounded). As the temperature decreases, the specific
heat increases up to a maximum value. The corresponding
temperature for this maximum value depends on the Fibonacci
generation number $n$, although one can see a clear tendency for a
common temperature value as $n$ increases. After the maximum
value, the specific heat falls into the low temperature region. In
this region it starts to present non-harmonic small oscillation
behavior, as shown in the inset of Fig.\ 2. These oscillations,
are {\it not} around a specific fractal dimension of the
quasiperiodic structure, as in the {\it idealized} triadic Cantor
set! Besides, it {\it cannot} be considered also an approximation
of the idealized oscillations found in the triadic Cantor set.
Their profiles define clearly two classes of oscillations, one for
the {\it even} and the other for the {\it odd} generation numbers
of the sequence, the amplitude of the odd oscillations being
bigger than the amplitude of the even one. These  behaviors are
better illustrated in Fig.\ 3, where are depicted {\it log-log}
plots of the specific heat against the temperature for several
generation numbers. Of course the number of oscillations observed
in the specific heat spectra is related to the hierarchical
generation number $n$ (more oscillations appear as $n$ increases).
Another behavior of these oscillations can be seen in Fig.\ 4,
where we have plotted the specific heat against $\log T$. We can
note a well defined period in the oscillations, which means that
the specific heat is a {\it log-periodic} function of the
temperature. The curves resemble the triadic case, with a mean
value $d$, around it $C(T)$ oscillates log-periodically, although,
as in the log-log plot, this value is not related with the fractal
dimension of the Fibonacci quasiperiodic structure. Besides, the
mean value is different for the even and odd Fibonacci's
generation numbers.

\section{Conclusions}

In this paper we have studied the spin wave specific heat, for
fixed values of the in-plane dimensionless wave vector $k_{x}a$,
for the Fibonacci magnetic superlattices. We have shown, as a
common aspect of the model employed here, that the specific heat
tends to zero in the high temperature limit ($T \rightarrow
\infty$),  as $T^{-2}$, no matter its generation number. This
asymptotic behavior is mainly due to the fact that we have
considered our system bounded. It also presents a maximum whose
corresponding temperature tends to a fixed value as the generation
number increases. Finally, in the low temperature region, a small
oscillations arises as an indication that the specific heat is a
{\it log-periodic} function of the temperature. These oscillations
can be defined as the {\it signature} of the quasiperiodic system,
and has no counterpart in the idealized case.

It would be of interest to have experimental data to test our
predicted theoretical results presented here. However, most of the
experimental studies of  magnetic multilayers to date, have been
done for transition metals \cite {16,17}. Further experimental
studies carried out for multilayers structures of magnetic
insulator or magnetic semiconductor materials, that would be
better described by the Heisenberg model employed in this paper,
are welcome.  Suitable experimental technique to probe the spin
wave spectra in the quasi-periodic structure discussed here is the
inelastic light scattering spectroscopy of Raman and Brillouin
type. However, the spectra can be obtained only for a given
in-plane wavevector $k_x$, which defines the incident angle
$\theta$ of the light through the relation
$\sin^{-1}\theta=k_x\lambda/4\pi$, $\lambda$ being the the laser
wavelength \cite {18,19}. This fact adds an extra complication to
the experimental specific heat measurements, but we hope that the
experimentalists can be encouraged to overcome it. Techniques
involving magnetic resonance (for example, ferromagnetic
resonance, standing spin-wave resonant etc.) can also be used, and
indeed they were previously been successfully applied to surface
and bulk spin waves in various magnetic microstructures (for a
good account of these techniques see \cite{20}).

\vskip 0.5cm

\noindent {\it Acknowledgments}: One of us (ELA) thanks the
hospitality of the Center for Polymer Studies, Boston
University-USA, where part of this work was done. We thank also
fruitful discussions with Profs. H.E. Stanley and S.V. Buldyrev,
as well as financial support from CNPq (Brazilian Agency) and
PRONEX.

\vskip 1.0 cm

\centerline {\bf Figure Captions}
 \vskip 1.0 cm

\begin{enumerate}

\item The spin wave spectra for the  Fibonacci  structure. Here
the in-plane wavevector is taken to be equal to 0.2.

\vskip 0.5cm

\item  Specific heat versus temperature for the Fibonacci  structure.
We have plotted the specific heat profiles up to the $9th$
generation number. The inset shows the low temperature behavior of
the specific heat.

\vskip 0.5cm

\item  Log-log plot of the specific heat versus temperature for the
generation numbers of the  Fibonacci quasiperiodic sequence.
Observe a different behavior for the even ($n$=4, 6, 8, and 10)
and odd ($n$=5, 7, and 9) generation numbers.

\vskip 0.5cm

\item  $C(T)$ vs $\log (T)$ plot to show the log-periodicity effects in
the Fibonacci structure.

\end{enumerate}

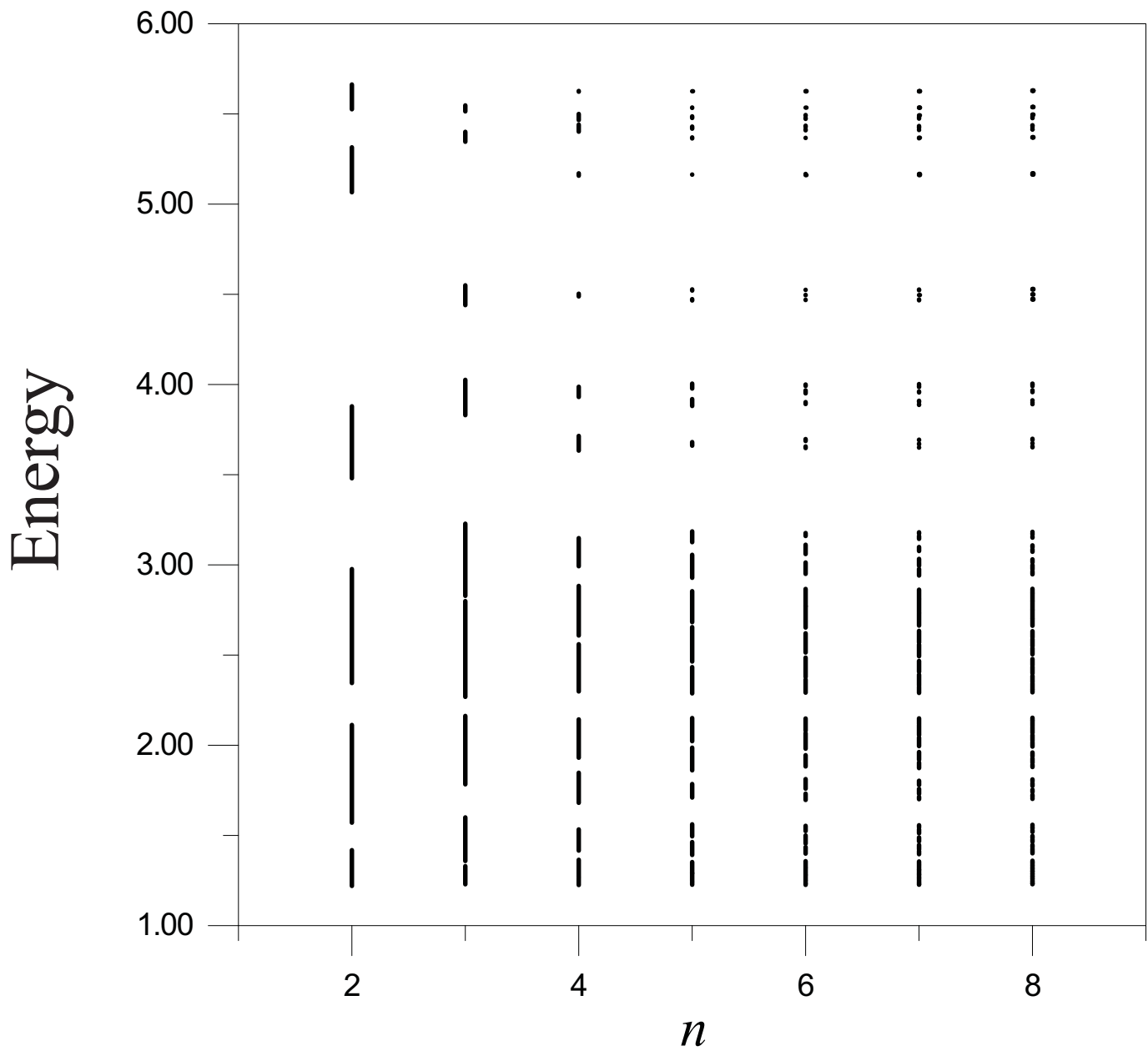
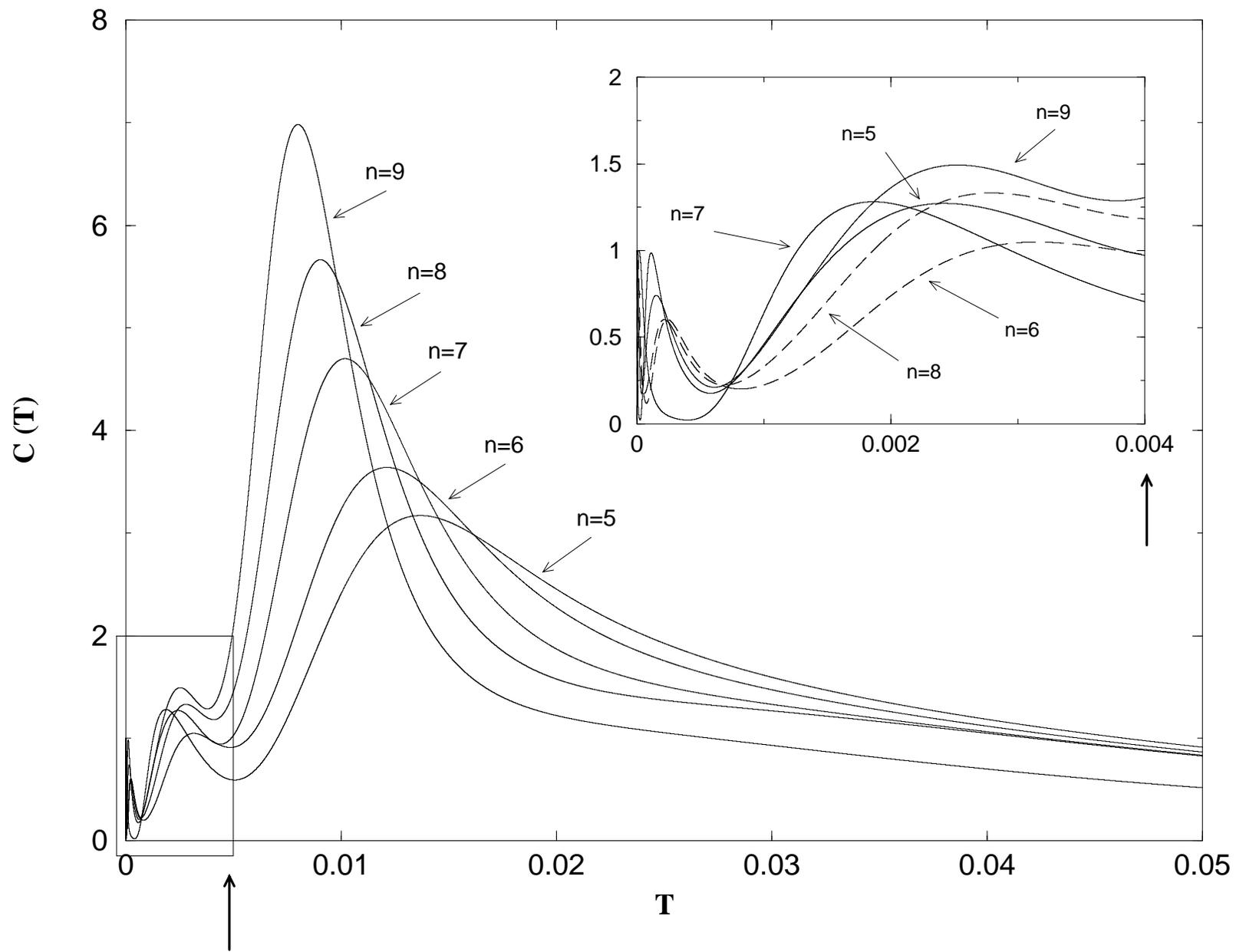
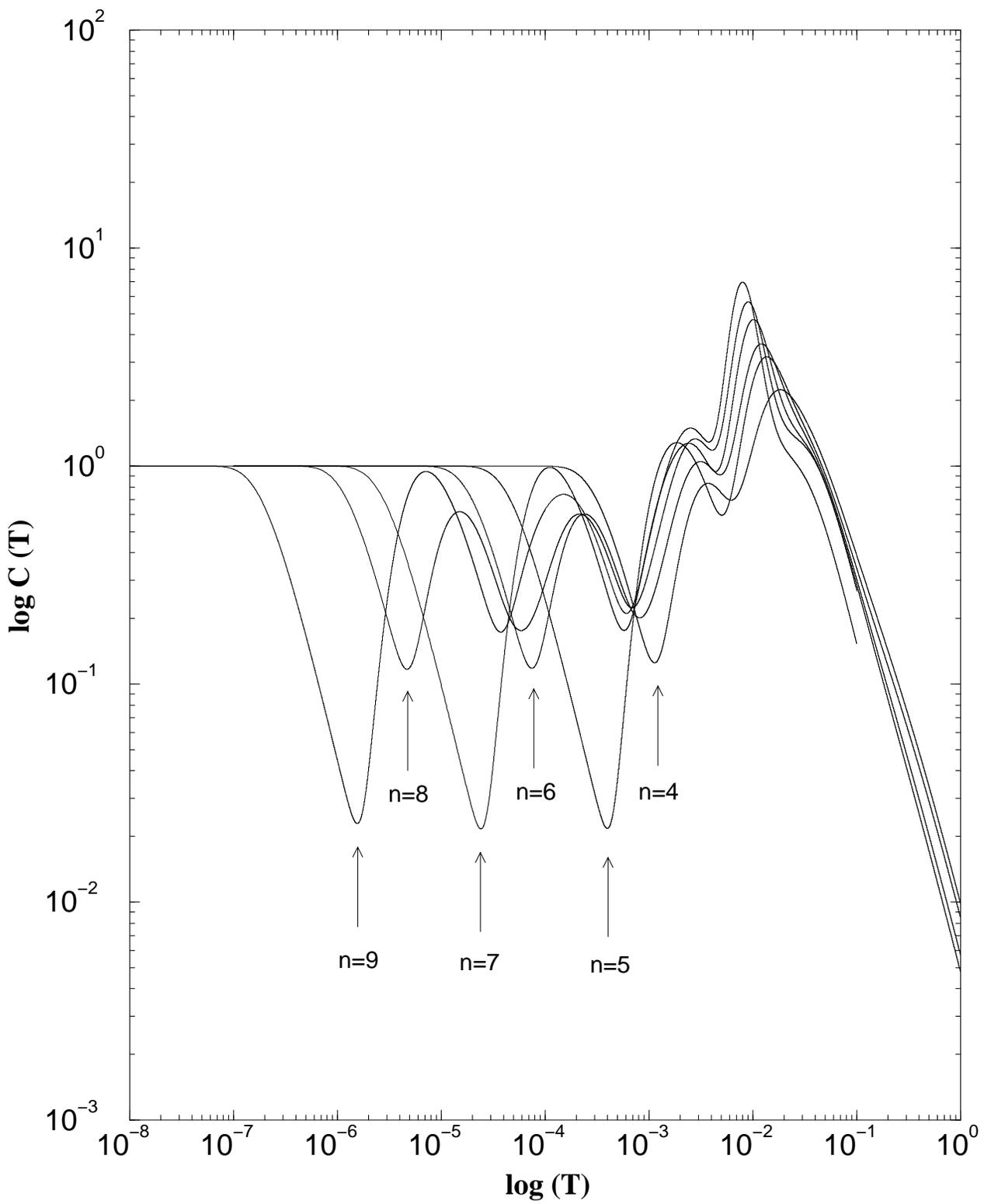
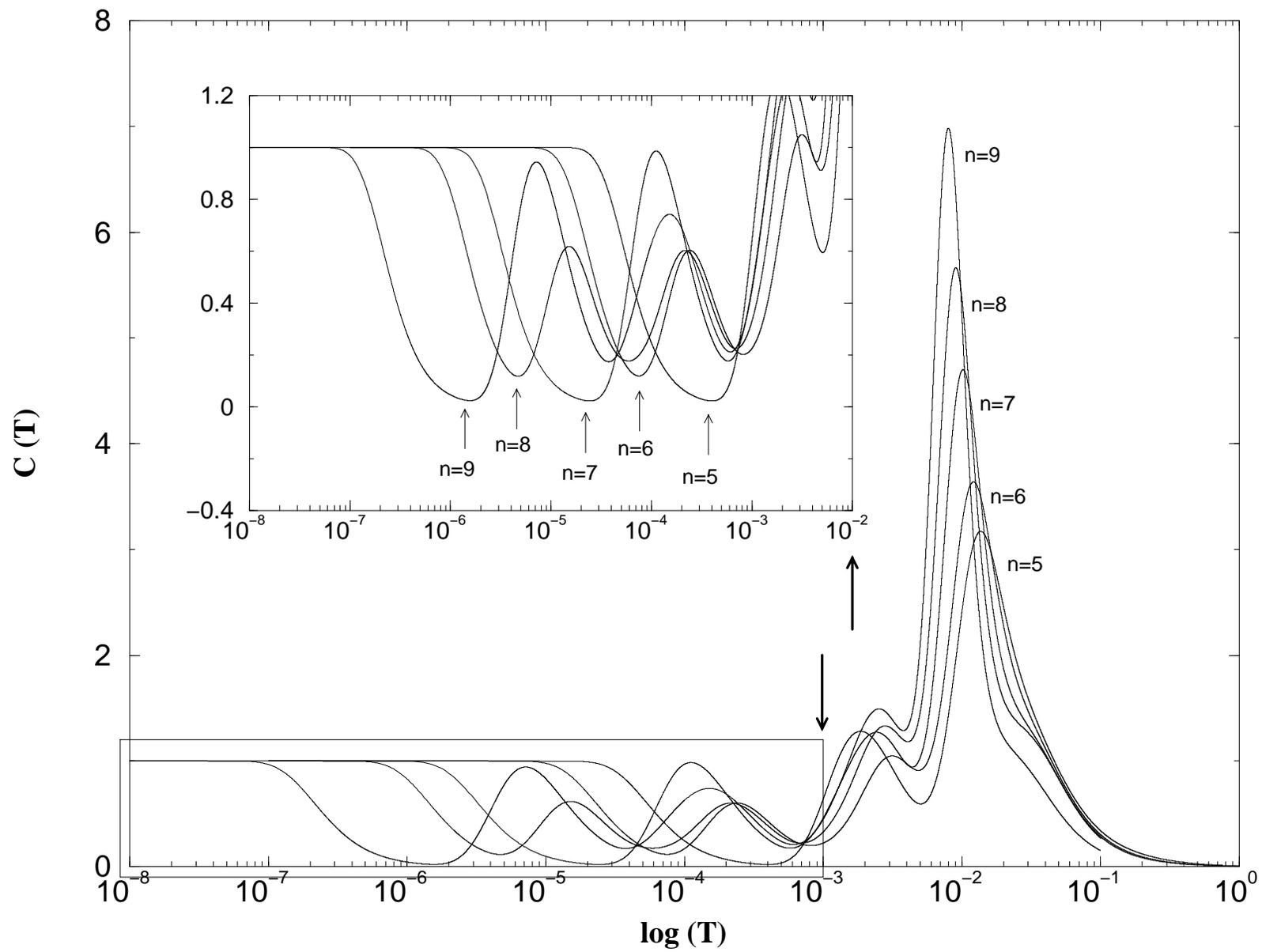
\end{document}